\documentclass[aps,pra,reprint,amsmath,amssymb,amsfonts,showkeys,showpacs,preprintnumbers,floatfix]{revtex4-1}
\usepackage{bm,graphicx,xcolor,lineno,hyperref,microtype,multirow,mathptmx}

\newcommand{\Hop}{\mathbf{H}}
\newcommand{\Ts}{T_{\rm S}}
\newcommand{\EJ}{E_{\rm J}}
\newcommand{\EK}{E_{\rm K}}
\newcommand{\EX}{E_{\rm X}}
\newcommand{\Ec}{E_{\rm C}}
\newcommand{\EV}{E_{\rm V}}
\newcommand{\ET}{E_{\rm T}}
\newcommand{\Eee}{E_{\rm ee}}
\newcommand{\EKS}{E_{\rm KS}}
\newcommand{\Ecjell}{E_{\rm C}^{\rm jell}}
\newcommand{\EcKS}{E_{\rm C}^{\rm KS}}
\newcommand{\EHF}{E_{\rm HF}}
\newcommand{\mc}{\multicolumn}
\newcommand{\md}{\mathcal{D}}
\newcommand{\br}{\mathbf{r}}

\newcommand{\rhor}{\rho(\mathbf{r})}

\begin{document}

\title{Uniform electron gases}

\author{Peter M. W. Gill}
\thanks{Corresponding author}
\email{peter.gill@anu.edu.au}
\affiliation{Research School of Chemistry, 
Australian National University, Canberra, ACT 0200, Australia}
\author{Pierre-Fran\c{c}ois Loos}
\email{loos@rsc.anu.edu.au}
\affiliation{Research School of Chemistry, 
Australian National University, Canberra, ACT 0200, Australia}
\date{\today}

\begin{abstract}
	We show that the traditional concept of the uniform electron gas (UEG) --- a homogeneous system of finite density, consisting of an infinite number of electrons in an infinite volume --- is inadequate to model the UEGs that arise in finite systems.  We argue that, in general, a UEG is characterized by at least two parameters, \textit{viz.}~the usual one-electron density parameter $\rho$ and a new two-electron parameter $\eta$.  We outline a systematic strategy to determine a new density functional $E(\rho,\eta)$ across the spectrum of possible $\rho$ and $\eta$ values.
\end{abstract}
\keywords{Uniform electron gas; Homogeneous electron gas; jellium; density functional theory}
\pacs{31.15.E-, 71.15.Mb, 71.10.Ca, 73.20.-r}

\maketitle

%========================
\section{Introduction} \label{intro}
%========================
The year 2012 is notable for both the journal and one of us for, in the months ahead, both TCA and PMWG will achieve their half-centuries.  It is inevitable and desirable that such occasions lead to retrospection, for it is often by looking backwards that we can perceive most clearly the way ahead.  Thus, as we ruminate on the things that we ought not to have done, we also dream of the things that we ought now to do.

The final decade of the 20th century witnessed a major revolution in quantum chemistry, as the subject progressed from an esoteric instrument of an erudite cognoscenti to a commonplace tool of the chemical proletariat.  The fuel for this revolution was the advent of density functional theory (DFT) \cite{ParrBook} models and software that were sufficiently accurate and user-friendly to save the experimental chemist some time.  These days, DFT so dominates the popular perception of molecular orbital calculations that many non-specialists now regard the two as synonymous.

In principle, DFT is founded in the Hohenberg-Kohn theorem \cite{HK64} but, in practice, much of its success can be traced to the similarity between the electron density in a molecule and the electron density in a hypothetical substance known as the uniform electron gas (UEG) or jellium \cite{Fermi26,Thomas27,Dirac30,Wigner34,Macke50,GellMann57,Onsager66,Stern73,Rajagopal77,Isihara80,Hoffman92,Seidl04,Sun10,2DEG11,3DEG11,VignaleBook}.  The idea --- the Local Density Approximation (LDA) --- is attractively simple:  if we know the properties of jellium, we can understand the electron cloud in a molecule by dividing it into tiny chunks of density and treating each as a piece of jellium.

The good news is that the properties of jellium are known from near-exact Quantum Monte Carlo calculations \cite{Ceperley80,Tanatar89,Kwon93,Ortiz94,Rapisarda96,Kwon98,Ortiz99,Attaccalite02,Zong02,Drummond09a,Drummond09b}.  Such calculations are possible because jellium is characterized by just a \emph{single} parameter $\rho$, the electron density.

The bad news is that jellium has an infinite number of electrons in an infinite volume and this unboundedness renders it, in some respects, a poor model for the electrons in molecules.  Indeed, the simple LDA described above predicts bond energies that are much too large and this led many chemists in the 1970s to dismiss DFT as a quantitatively worthless theory.

Most of the progress since those dark days has resulted from concocting ingenious corrections for jellium's deficiencies.  For example, significant improvements in accuracy can be achieved by using both the density $\rhor$ and its gradient $\nabla\rhor$, an approach called gradient-corrected DFT \cite{Perdew86}.  Even better results can be achieved by including a fraction of Hartree-Fock exchange (yielding hybrid methods \cite{Hybrid92,Becke93}) or higher derivatives of $\rhor$ (leading to meta-GGAs \cite{Tau03}).

However, notwithstanding the impressive progress since the 1970s, modern DFT approximations still exhibit fundamental deficiencies in large systems \cite{Curtiss00}, conjugated molecules \cite{Woodcock02}, charge-transfer excited states \cite{Dreuw04}, dispersion-stabilized systems \cite{Wodrich06}, systems with fractional spin or charge \cite{Yang08}, isodesmic reactions \cite{DeFecT09} and elsewhere.  Because DFT is in principle an exact theory, many of these problems can be traced ultimately to the use of jellium as a reference system and the \textit{ad hoc} corrections that its use subsequently necessitates.  It is not a good idea to build one's house on sand!

In an attempt to avoid some of the weaknesses of jellium-based DFT, we have invented and explored an alternative paradigm called Intracule Functional Theory (IFT) \cite{Overview03,Omega06,AnnuRep11}.  In this approach, the one-electron density $\rhor$ is abandoned in favour of two-electron variables (such as the interelectronic distance $r_{12}$) and we have discovered that the latter offer an efficient and accurate route to the calculation of molecular energies \cite{Sin3w07,AnalKer07,OmegaComp07,Dot08,BasisEffects09,RR11}.  Nonetheless, IFT is not perfect and has shortcomings that are complementary to those of DFT.  As a result, one should seek to combine the best features of each, to obtain an approach superior to both.  That is the goal of the present work and we will use atomic units throughout this article.

%================================
\section{Electrons on spheres} \label{sec2}
%================================
In recent research, we were led to consider the behavior of electrons that are confined to the surface of a ball.  This work yielded a number of unexpected discoveries \cite{TEOAS09,EcLimit09,QuasiExact09,TEOCS09,LoosHL10,EcProof10,ExSpherium10,Frontiers10,Glomium11} but the one of relevance here is that such systems provide a beautiful new family of uniform electron gases (see also \cite{Seidl07}).

\begin{table}
	\caption{The lowest free-particle orbitals on a 2-sphere}
	\label{tab:Ylm}
	\begin{tabular}{c@{\qquad}cc@{\qquad}c}
	\hline\noalign{\smallskip}
		Name		&	$l$	&	$m$	&	$\sqrt{4\pi} \,Y_{lm}(\theta,\phi)$							\\
	\noalign{\smallskip}\hline\noalign{\smallskip}
		$s$			&	0	&	\ \,0		&	$1$															\\
	\noalign{\smallskip}\hline\noalign{\smallskip}
		$p_0$		&	1	&	\ \,0		&	$3^{1/2} \cos\theta$										\\
		$p_{+1}$	&	1	&	+1		&	$(3/2)^{1/2} \sin\theta \exp(+i\phi)$							\\
		$p_{-1}$	&	1	&	--1		&	$(3/2)^{1/2} \sin\theta \exp(-i\phi)$							\\
	\noalign{\smallskip}\hline\noalign{\smallskip}
		$d_0$		&	2	&	\ \,0		&	$(5/4)^{1/2} (3\cos^2\theta-1)$								\\
		$d_{+1}$	&	2	&	+1		&	$(15/2)^{1/2} \sin\theta \cos\theta \exp(+i\phi)$				\\
		$d_{-1}$	&	2	&	--1		&	$(15/2)^{1/2} \sin\theta \cos\theta \exp(-i\phi)$				\\
		$d_{+2}$	&	2	&	+2		&	$(15/8)^{1/2} \sin^2\theta \exp(+2i\phi)$					\\
		$d_{-2}$	&	2	&	--2		&	$(15/8)^{1/2} \sin^2\theta \exp(-2i\phi)$						\\
	\noalign{\smallskip}\hline
	\end{tabular}
\end{table}

\begin{table}
	\caption{Number of electrons in $L$-spherium and $L$-glomium atoms}
	\label{tab:fullshell}
	\begin{tabular}{ccccccccc}
	\hline\noalign{\smallskip}
		$L$				&	0	&	1	&	2	&	3	&	4		&	5		&	6		&	7		\\
	\noalign{\smallskip}\hline\noalign{\smallskip}
		$L$-spherium	&	2	&	8	&	18	&	32	&	50		&	72		&	98		&	128		\\
		$L$-glomium	&	2	&	10	&	28	&	60	&	110		&	182		&	280		&	408		\\
	\noalign{\smallskip}\hline
	\end{tabular}
\end{table}

%-----------------------------------------------------------------------
\subsection{Spherium atoms} \label{subsec:spherium}
%-----------------------------------------------------------------------
The surface of a three-dimensional ball is called a 2-sphere (for it is two-dimensional) and its free-particle orbitals (Table \ref{tab:Ylm}) are the spherical harmonics $Y_{lm}(\theta,\phi)$.  It is known that
\begin{equation}
	\sum_{m=-l}^l |Y_{lm}(\theta,\phi) |^2 = \frac{2l+1}{4\pi}
\end{equation}
and doubly occupying all the orbitals with $0 \le l \le L$ thus yields a uniform electron gas.  We call this system $L$-spherium and will compare it to \emph{two}-dimensional jellium \cite{VignaleBook}.

The number of electrons (Table \ref{tab:fullshell}) in $L$-spherium is
\begin{equation}
	n = 2(L+1)^2
\end{equation}
the volume of a 2-sphere is $V = 4\pi R^2$ and, therefore,
\begin{equation} \label{eq:rho2d}
	\rho = \frac{(L+1)^2}{2\pi R^2}
\end{equation}

\begin{table}
	\caption{The lowest free-particle orbitals on a glome (\textit{i.e.}~a 3-sphere)}
	\label{tab:Ylmn}
	\begin{tabular}{c@{\qquad}ccc@{\qquad}c}
	\hline\noalign{\smallskip}
		Name		&	$l$	&	$m$&	$n$		&	$\pi\, Y_{lmn}(\chi,\theta,\phi)$							\\
	\noalign{\smallskip}\hline\noalign{\smallskip}
		$1s$		&	0	&	0	&	\ \,0		&	$2^{-1/2}$												\\
	\noalign{\smallskip}\hline\noalign{\smallskip}
		$2s$		&	1	&	0	&	\ \,0		&	$2^{1/2} \cos\chi$										\\
		$2p_0$		&	1	&	1	&	\ \,0		&	$2^{1/2} \sin\chi \cos\theta$							\\
		$2p_{+1}$	&	1	&	1	&	+1		&	$\sin\chi \sin\theta \exp(+i\phi)$							\\
		$2p_{-1}$	&	1	&	1	&	--1		&	$\sin\chi \sin\theta \exp(-i\phi)$							\\
	\noalign{\smallskip}\hline\noalign{\smallskip}
		$3s$		&	2	&	0	&	\ \,0		&	$2^{-1/2} (4\cos^2\chi-1)$								\\
		$3p_0$		&	2	&	1	&	\ \,0		&	$12^{1/2}\sin\chi \cos\chi \cos\theta$					\\
		$3p_{+1}$	&	2	&	1	&	+1		&	$6^{1/2}\sin\chi \cos\chi \sin\theta \exp(+i\phi)$			\\
		$3p_{-1}$	&	2	&	1	&	--1		&	$6^{1/2}\sin\chi \cos\chi \sin\theta \exp(-i\phi)$			\\
		$3d_0$		&	2	&	2	&	\ \,0		&	$\sin^2\chi \ (3\cos^2\theta-1)$							\\
		$3d_{+1}$	&	2	&	2	&	+1		&	$6^{1/2}\sin^2\chi \sin\theta \cos\theta \exp(+i\phi)$	\\
		$3d_{-1}$	&	2	&	2	&	--1		&	$6^{1/2}\sin^2\chi \sin\theta \cos\theta \exp(-i\phi)$		\\
		$3d_{+2}$	&	2	&	2	&	+2		&	$(3/2)^{1/2}\sin^2\chi \sin^2\theta \exp(+2i\phi)$		\\
		$3d_{-2}$	&	2	&	2	&	--2		&	$(3/2)^{1/2}\sin^2\chi \sin^2\theta \exp(-2i\phi)$		\\
	\noalign{\smallskip}\hline
	\end{tabular}
\end{table}

%---------------------------------------------------------------------
\subsection{Glomium atoms} \label{subsec:glomium}
%---------------------------------------------------------------------
The surface of a four-dimensional ball is a 3-sphere (or ``glome'' \cite{glome}) and its free-particle orbitals (Table \ref{tab:Ylmn}) are the hyperspherical harmonics $Y_{lmn}(\chi,\theta,\phi)$.  It is known \cite{AveryBook} that
\begin{equation}
	\sum_{m=0}^l \sum_{n=-m}^m |Y_{lmn}(\chi,\theta,\phi) |^2 = \frac{(l+1)^2}{2\pi^2}
\end{equation}
and doubly occupying all the orbitals with $0 \le l \le L$ thus yields a uniform electron gas.  We call this system $L$-glomium and will compare it to \emph{three}-dimensional jellium \cite{VignaleBook}.

The number of electrons (Table \ref{tab:fullshell}) in $L$-glomium is
\begin{equation}
	n = (L+1)(L+2)(2L+3)/3
\end{equation}
the volume of a 3-sphere is $V = 2\pi^2 R^3$ and, therefore,
\begin{equation}
	\rho = \frac{(L+1)(L+2)(2L+3)}{6\pi^2 R^3}
\end{equation}

%-----------------------------------------------------------------------------
\subsection{Exactly solvable systems} \label{subsec:quasi}
%-----------------------------------------------------------------------------
One of the most exciting features of the two-electron atoms 0-spherium and 0-glomium is that, for certain values of the radius $R$, their Schr\"odinger equations are exactly solvable \cite{QuasiExact09}.  The basic theory is as follows.

The Hamiltonian for two electrons on a sphere is
\begin{equation}
	\mathbf{H} = - \frac{\nabla_1^2}{2} - \frac{\nabla_2^2}{2} + \frac{1}{u}
\end{equation}
where $u$ is the interelectronic distance $r_{12} \equiv |\br_1-\br_2|$.  If we assume that the Hamiltonian possesses eigenfunctions that depend only on $u$, it becomes
\begin{equation}
	\mathbf{H} = \left[ \frac{u^2}{4R^2} - 1 \right] \frac{d^2}{du^2} + \left[ \frac{(2\md-1)u}{4R^2} - \frac{\md-1}{u} \right] \frac{d}{du} + \frac{1}{u}
\end{equation}
where $\md$ is the dimensionality of the sphere.  Three years ago, we discovered that $\mathbf{H}$ has polynomial eigenfunctions, but only for particular values of $R$.  (This is analogous to the discovery that hookium\footnote{The hookium atom consists of two electrons that repel Coulombically but are bound to the origin by a harmonic potential \cite{Kestner62}.} has closed-form wavefunctions, but only for particular harmonic force constants \cite{Kais89,Taut93}.)  We showed that there exist $\lfloor (n+1)/2 \rfloor$ $n$th-degree polynomials of this type and that the associated energies and radii satisfy
\begin{equation} \label{eq:4R2E}
	4R^2_{n,m} E_{n,m} = n(n+2\md-2)
\end{equation}
where the index $m = 1, \ldots , \lfloor (n+1)/2 \rfloor$.

For 0-spherium (\textit{i.e.}~$\md=2$), by introducing $x = u/(2R)$ and using Eq.~(\ref{eq:4R2E}), we obtain the Sturm-Liouville equation
\begin{equation}
	\frac{d}{dx} \left[ \frac{x(1-x^2)}{2} \frac{d\Psi}{dx} \right] + \frac{n(n+2)x}{2} \Psi = R \Psi
\end{equation}
The eigenradii $R$ can then be found by diagonalization in a polynomial basis which is orthogonal on $[0,1]$.  The shifted Legendre polynomials are ideal for this \cite{NISTbook}.

For 0-glomium (\textit{i.e.}~$\md=3$), proceeding similarly yields
\begin{equation}
	\frac{d}{dx} \left[ \frac{x(1-x^2)}{2} w(x) \frac{d\Psi}{dx} \right] + \frac{n(n+4)x}{2} w(x) \Psi = R w(x) \Psi
\end{equation}
where the weight function $w(x) = x\sqrt{1-x^2}$.  The eigenradii are found by diagonalization in a basis which is orthogonal with respect to $w(x)$ on $[0,1]$.

An exact energy can be partitioned into its kinetic part
\begin{equation}
	\ET = (-1/4) \langle \Psi | \nabla_1^2 + \nabla_2^2 | \Psi \rangle
\end{equation}
and its two-electron part
\begin{equation}
	\Eee = (1/2) \langle \Psi | u^{-1} | \Psi \rangle
\end{equation}
and the resulting reduced energies (\textit{i.e.}~the energy per electron) of the ground states of 0-spherium and 0-glomium, for the first two eigenradii, are shown in the left half of Table \ref{tab:properties}.

%============================================
\section{Single-determinant methods} \label{sec:HFandKS}
%============================================
%--------------------------------------------------------------------------
\subsection{Hartree-Fock theory \cite{Hartree28,Fock30}}
%--------------------------------------------------------------------------
In the Hartree-Fock (HF) partition, the reduced energy\footnote{Henceforth, all energies will be reduced energies.} of an $n$-electron system is
\begin{equation}  \label{eq:EHF}
	E = \Ts + \EV + \EJ + \EK + \Ec
\end{equation}
where the five contributions are the non-interacting-kinetic, external, Hartree, exchange and correlation energies, respectively.  The first four of these are defined by
\begin{gather}
	\Ts = -\frac{1}{2n} \sum_i^n \int \psi_i^*(\br) \nabla^2 \psi_i(\br) \,d\br		\\
	\EV = + \frac{1}{n} \int \rhor v(\br) \,d\br										\\
	\EJ = + \frac{1}{2n} \iint \rho(\br_1) r_{12}^{-1} \rho(\br_2) \,d\br_1 \,d\br_2	\\
	\EK = - \frac{1}{2n} \sum_{i,j}^n \iint \psi_i^*(\br_1)\psi_j(\br_1) r_{12}^{-1} \psi_j^*(\br_2) \psi_i(\br_2) \,d\br_1 \,d\br_2
\end{gather}
where $\psi_i(\br)$ is an occupied orbital, $\rhor$ is the electron density, and $v(\br)$ is the external potential.  The correlation energy $\Ec$ is defined so that Eq.~(\ref{eq:EHF}) is exact.

%---------------------------------------------------------------------------------
\subsection{Kohn-Sham density functional theory \cite{KS65}}
%---------------------------------------------------------------------------------
In the Kohn-Sham (KS) partition, the energy is
\begin{equation} \label{eq:EKS}
	\EKS = \Ts + \EV + \EJ + \EX + \EcKS
\end{equation}
where the last two terms, which are sometimes combined, are the Kohn-Sham exchange and correlation energies.  The correlation energy $\EcKS$ is defined so that Eq.~(\ref{eq:EKS}) is exact.

Many formulae have been proposed for $\EX$ and $\EcKS$ but the most famous are those explicitly designed to be exact for $\md$-jellium.  In the case of exchange, one finds
\begin{equation} \label{eq:Ex}
	\EX = X_\md \int \rho^{1/\md} \,d\br
\end{equation}
where Dirac \cite{Dirac30} determined the coefficient $X_3$ in 1930 and Glasser \cite{Glasser83} found the general formula for $X_\md$ in 1983.  The correlation functional is not known exactly but accurate Quantum Monte Carlo calculations on jellium in 2D \cite{Tanatar89,Kwon93,Ortiz94,Rapisarda96,Ortiz99,Attaccalite02,Drummond09a,Drummond09b} and 3D \cite{Ceperley80,Kwon98,Zong02} have been fitted \cite{Attaccalite02,Perdew92} to functions of the form
\begin{equation} \label{eq:Ec}
	\Ecjell = \int C_\md(\rho) \,d\br
\end{equation}

By construction, Eq.~(\ref{eq:Ec}) yields the correct energy when applied to the uniform electron gas in jellium.  But what happens when we apply it to a uniform electron gas on a sphere?

\begin{table*}
	\caption{Exact and Kohn-Sham reduced energies of the ground states of 0-spherium and 0-glomium for various eigenradii $R$}
	\label{tab:properties}
	\begin{tabular}{ccccclcccccclc}
					\hline\noalign{\smallskip}
				&				&			\mc{3}{c}{Exact}			&&				\mc{6}{c}{Jellium-based Kohn-Sham DFT}				&&	Error	\\
					\noalign{\smallskip}\cline{3-5}\cline{7-12}\noalign{\smallskip}
				&	$2R$		&	$\ET$		&	$\Eee$		&  $E$	&&$\Ts$&$\EV$&	$\EJ$		&	$-\EX$		& $-\Ecjell$	&  $\EKS$	&&$\EKS-E$\\
					\noalign{\smallskip}\hline\noalign{\smallskip}
	0-spherium	&  $\sqrt{3}$	&	0.051982	&	0.448018	&	1/2	&&	0	&	0	&	1.154701	&	0.490070	&	0.1028	&	0.562	&&	0.062	\\
				&  $\sqrt{28}$	&	0.018594	&	0.124263	&	1/7	&&	0	&	0	&	0.377964	&	0.160413	&	0.0593	&	0.158	&&	0.015	\\
					\noalign{\smallskip} \noalign{\smallskip}
	0-glomium	&  $\sqrt{10}$	&	0.014213	&	0.235787	&	1/4	&&	0	&	0	&	0.536845	&	0.217762	&	0.0437	&	0.275	&&	0.025	\\
				&  $\sqrt{66}$	&	0.007772	&	0.083137	&  1/11	&&	0	&	0	&	0.208967	&	0.084764	&	0.0270	&	0.097	&&	0.006	\\
					\noalign{\smallskip}\hline
	\end{tabular}
\end{table*}

%============================================
\section{The non-uniqueness problem} \label{sec:problem}
%============================================
The deeply disturbing aspect of jellium-based DFT models --- and the launching-pad for the remainder of this paper --- is the countercultural claim, that
\begin{center}
	\emph{The uniform electron gas with density $\rho$ is not unique.}
\end{center}
Though it may seem heretical to someone who has worked with jellium for many years, or to someone who suspects that the claim violates the Hohenberg-Kohn theorem, we claim that two $\md$-dimensional uniform electron gases with the same density parameter $\rho$ may have different energies.  To illustrate this, we now show that density functionals \cite{Dirac30,Perdew92,Attaccalite02} which are exact for jellium are wrong for 0-spherium and 0-glomium.

%-----------------------------------------------------------------------
\subsection{Illustrations from exactly solvable systems}
%-----------------------------------------------------------------------
The energy contributions for 0-spherium and 0-glomium are easy to find.  There is no external potential, so $\EV = 0$.  The density $\rhor$ is constant, so the Kohn-Sham orbital $\psi(\br) = \sqrt{\rhor}$ is constant, and $\Ts=0$.  The Hartree energy is the self-repulsion of a uniform spherical shell of charge of radius $R$ and one finds \cite{EcLimit09}
\begin{equation}
	\EJ = \frac{\Gamma(\md-1)}{\Gamma(\md-1/2)} \frac{\Gamma(\md/2+1/2)}{ \Gamma(\md/2)} \frac{1}{R}
\end{equation}
The exchange energy is predicted \cite{Glomium11} by Eq.~(\ref{eq:Ex}) to be
\begin{equation}
	\EX = - \frac{2\md}{(\md^2-1)\pi R} \left(\frac{\md!}{2} \right)^{1/\md}
\end{equation}
and the correlation energy predicted by Eq.~(\ref{eq:Ec}) is simply
\begin{equation}
	\Ecjell = C_\md(2/V)
\end{equation}
Applying these formulae to the exactly solvable states of 0-spherium and 0-glomium considered in Section \ref{subsec:quasi} yields the results in the right half of Table \ref{tab:properties}.  In all cases, the KS-DFT energies are too high by 10 -- 20\%, indicating that the correlation functional that is exact for the uniform electron gas in jellium grossly underestimates the correlation energy of the uniform electron gases in 0-spherium and 0-glomium.

%-------------------------------------------------------------------------------------------
\subsection{Limitations of the one-electron density parameter $\rho$}
%-------------------------------------------------------------------------------------------
The results in Table \ref{tab:properties} demonstrate conclusively that not all uniform electron gases with the density $\rho$ are equivalent.  The simplest example of this is the ground state of two electrons on a 2-sphere with $R = \sqrt{3}/2$.  The exact wavefunction, reduced energy and density of this system are
\begin{gather}
	\Psi = 1+u			\\
	E = 1/2				\\
	\rho(\br) = 2 / (3\pi)
\end{gather}
but, when fed this uniform density, the exchange-correlation functional that is exact for two-dimensional jellium grossly overestimates the energy, yielding $\EKS = 0.562$.

This discovery has worrying chemical implications.  Contrary to the widespread belief that the LDA (\textit{e.g.}~the S-VWN functional) is accurate when applied to regions of a molecule where $\rhor$ is almost uniform (such as near bond midpoints), our results reveal that it actually performs rather poorly.

The discovery also has counterintuitive implications at a theoretical level.  It implies that the years of effort that have been expended in calculating the properties of jellium do \emph{not} provide us with a complete picture of homogeneous electron gases.  On the contrary, although they inform us in detail about the \emph{infinite} uniform electron gas, they tell us very little about the properties of finite electron gases.

In a nutshell, the results in Table \ref{tab:properties} reveal that a UEG is not completely characterized by its one-electron density parameter $\rho$.  Evidently, something else is required\ldots

%------------------------------------------------------------------------
\subsection{Virtues of two-electron density parameters}
%------------------------------------------------------------------------
We know that it is possible for two uniform electron gases to have the same density $\rho$ but different reduced energies $E$.  But how can this be, given that the probability of finding an electron in a given volume is identical in the two systems?  The key insight is that the probability of finding \emph{two} electrons in that volume is different.

\begin{figure}
	\includegraphics[width=0.48\textwidth]{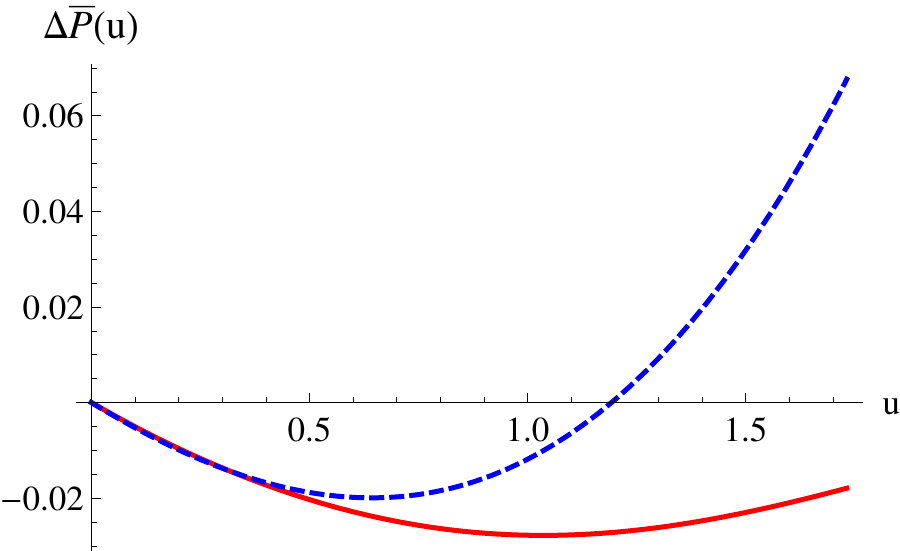}
	\caption{Specific Coulomb holes for 0-spherium (dotted) with $R=\sqrt{3}/2$ and 2D jellium (solid).  Both are uniform gases with $\rho = 2/(3\pi)$.}
	\label{fig:PuR1}
\end{figure}

\begin{figure}
	\includegraphics[width=0.48\textwidth]{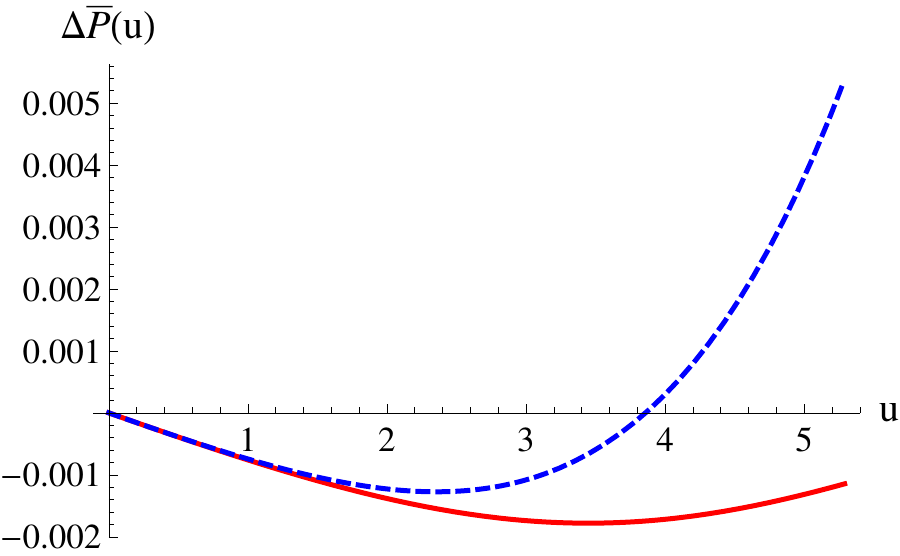}
	\caption{Specific Coulomb holes for 0-spherium (dotted) with $R=\sqrt{7}$ and 2D jellium (solid).  Both are uniform gases with $\rho = 1/(14\pi)$.}
	\label{fig:PuR2}
\end{figure}

This is illustrated in Figs \ref{fig:PuR1} and \ref{fig:PuR2}, which compare the probability distributions of the interelectronic distance \cite{Coulson61,GoriGiorgi04,TEOCS09} in various two-dimensional uniform electron gases.  These reveal that, although similar for $u \approx 0$ (because of the Kato cusp condition \cite{Kato57}), the specific Coulomb holes (\textit{i.e.} the holes per unit volume \cite{Overview03}) in two gases with the same one-electron density $\rho$ can be strikingly different.  In each case, the jellium hole is both deeper and wider than the corresponding spherium hole, indicating that the jellium electrons exclude one another more strongly, and one is much less likely to find two electrons in a given small volume of jellium than in the same volume of spherium.

We conclude from these comparisons that (at least) two parameters are required to characterize a uniform electron gas.  Although the parameter choice is not unique, we believe that the first should be a one-electron quantity, such as the density $\rho$ (or, equivalently, the Seitz radius $r_s$) and the second should be a two-electron quantity such as $\eta = h(\br,\br)$, where $h$ is the pair correlation function defined by
\begin{equation}
	\rho_2(\br_1,\br_2) = \frac{1}{2} \rho(\br_1) \rho(\br_2) \left[ 1 + h(\br_1,\br_2) \right]
\end{equation}
and $\rho_2$ is the diagonal part of the second-order spinless density matrix \cite{ParrBook}.

%===================================
\section{Lessons from Spherium and Glomium}
%===================================
%----------------------------------------
\subsection{A modest proposal}
%----------------------------------------
The discovery that uniform electron gases are characterized by \emph{two} parameters ($\rho$ and $\eta$) has many ramifications but one of the most obvious is that the foundations of the venerable Local Density Approximation need to be rebuilt.

The traditional LDA writes the correlation energy of a molecular system as
\begin{equation} \label{eq:oldlda}
	\EcKS \approx \int C(\rho) \,d\br
\end{equation}
thereby assuming that the contribution from each point $\br$ depends only on the one-electron density $\rhor$ at that point.  However, now that we know that the energy of a uniform electron gas depends on $\rho$ and $\eta$, it is natural to replace Eq.~(\ref{eq:oldlda}) by the generalized expression
\begin{equation} \label{eq:newlda}
	\EcKS \approx \int C(\rho,\eta) \,d\br
\end{equation}
In a sense, this two-parameter LDA represents a convergence in the evolution of Density Functional Theory (which stresses the one-electron density) and Intracule Functional Theory (which focuses on the two-electron density).

How can we find the new density functional $C(\rho,\eta)$?  One could take an empirical approach but that is an overused option within the DFT community \cite{Obituary01} and we feel that it is more satisfactory to derive it from the uniform electron gas.  As we show in Section \ref{subsec:HFcalcs}, it turns out that it is easy to compute the exact values of $\Ts$, $\EV$, $\EJ$ and $\EX$ for any $L$-spherium or $L$-glomium atom and, therefore, if one knew the exact wavefunctions and energies of $L$-spherium and $L$-glomium for a wide range of $L$ and $R$, one could extract the exact Kohn-Sham correlation energies
\begin{equation}
	\EcKS = E - \Ts - \EV - \EJ - \EX
\end{equation}
and determine the exact dependence of these on $\rho$ and $\eta$.

Accordingly, we propose to embark on a comprehensive study of $L$-spherium and $L$-glomium atoms, in order eventually to liberate the LDA from jellium's yoke through a process of radical generalization.  The results of these spherium and glomium calculations will generalize the known properties of jellium, because we have shown that the energies of $L$-spherium and $L$-glomium approach those of 2D jellium and 3D jellium, as $L$ becomes large.

In the remaining sections, we will confine our attention to the (two-dimensional) $L$-spherium atoms.  However, exactly the same approach can and will be used to address the (three-dimensional) $L$-glomium atoms in the future.

\begin{table*}
	\caption{Reduced energies, densities and $\eta$ values of $L$-spherium with the four smallest eigenradii$^\dagger$ $R$}
	\label{tab:results2}
	\begin{tabular}{ccrrrlrrrlrrr}
			\hline\noalign{\smallskip}
		&			&	\mc{3}{c}{Wavefunction-based energies}	&&		\mc{3}{c}{Kohn-Sham energies}		&&	\mc{3}{c}{Ingredients of the new model}	\\
			\noalign{\smallskip}\cline{3-5}\cline{7-9}\cline{11-13}\noalign{\smallskip}
	$L$	&	$R$	&$\EHF\quad$&  $-\Ec\quad$	&	$E\quad$	&&	$\Ts\quad$	&  $\EJ\quad$	&  $-\EX\quad$	&&$-\EcKS\ \ \ \ $&$\rho\quad$	& $\eta\quad$	\\
			\noalign{\smallskip}\hline\noalign{\smallskip}
	0	&	$R_1$	&	0.577350	&	0.077350	&	0.500000	&&	0.000000	&	1.154701	&	0.490070	&&	0.164630	&	0.212207	&	-0.896037	\\
		&	$R_2$	&	0.188982	&	0.046125	&	0.142857	&&	0.000000	&	0.377964	&	0.160413	&&	0.074694	&	0.022736	&	-0.991159\\
		&	$R_3$	&	0.092061	&	0.028497	&	0.063564	&&	0.000000	&	0.184122	&	0.078144	&&	0.042414	&	0.005396	&	-0.999496	\\
		&	$R_4$	&	0.054224	&	0.018941	&	0.035282	&&	0.000000	&	0.108447	&	0.046026	&&	0.027138	&	0.001872	&	-0.999976	\\
			\noalign{\smallskip} \noalign{\smallskip}
	1	&	$R_1$	&	4.579572	&				&				&&	1.000000	&	4.618802	&	0.980140	&&				&	0.848826	&				\\
		&	$R_2$	&	1.278833	&				&				&&	0.107143	&	1.511858	&	0.320826	&&				&	0.090946	&				\\
		&	$R_3$	&	0.596204	&				&				&&	0.025426	&	0.736488	&	0.156288	&&				&	0.021582	&				\\
		&	$R_4$	&	0.345007	&				&				&&	0.008821	&	0.433789	&	0.092053	&&				&	0.007487	&				\\
			\noalign{\smallskip} \noalign{\smallskip}
	2	&	$R_1$	&	11.543198	&				&				&&	2.666667	&	10.392305	&	1.470210	&&				&	1.909859	&				\\
		&	$R_2$	&	3.191241	&				&				&&	0.285714	&	3.401680	&	0.481239	&&				&	0.204628	&				\\
		&	$R_3$	&	1.483203	&				&				&&	0.067802	&	1.657098	&	0.234431	&&				&	0.048560	&				\\
		&	$R_4$	&	0.857188	&				&				&&	0.023522	&	0.976025	&	0.138079	&&				&	0.016846	&				\\
			\noalign{\smallskip} \noalign{\smallskip}
	3	&	$R_1$	&	21.477457	&				&				&&	5.000000	&	18.475209	&	1.960281	&&				&	3.395305	&				\\
		&	$R_2$	&	5.929228	&				&				&&	0.535714	&	6.047432	&	0.641652	&&				&	0.363783	&				\\
		&	$R_3$	&	2.754531	&				&				&&	0.127128	&	2.945952	&	0.312575	&&				&	0.086328	&				\\
		&	$R_4$	&	1.591634	&				&				&&	0.044103	&	1.735156	&	0.184106	&&				&	0.029949	&				\\
			\noalign{\smallskip}\hline
	\end{tabular}
	\vspace{1mm} \\
	$^\dagger$ $R_1 = \frac{1}{2}\sqrt{3}$; \ $R_2 = \frac{1}{2}\sqrt{28}$; \ $R_3 = \frac{1}{2}\sqrt{63+12\sqrt{21}}$; \ $R_4 = \frac{1}{2}\sqrt{198+6\sqrt{561}}$
\end{table*}

%-----------------------------------------------
\subsection{Basis sets and integrals}
%-----------------------------------------------
The Hamiltonian for $L$-spherium is
\begin{equation}
	\Hop = - \frac{1}{2} \sum_{i=1}^n \nabla_i^2 + \sum_{i<j}^n \frac{1}{r_{ij}}
\end{equation}
and the natural basis functions for HF and correlated calculations on this are the spherical harmonics $Y_{lm}(\theta,\phi)$ introduced in Section \ref{sec2}.  These functions are orthonormal \cite{NISTbook}
\begin{equation}
	\big\langle Y_{lm} \big| Y_{l'm'} \big\rangle = \delta_{l,l'} \ \delta_{m,m'}
\end{equation}
and are eigenfunctions of the Laplacian, so that
\begin{equation} \label{eq:kinint}
	\big\langle Y_{lm} \big| \nabla^2 \big| Y_{l'm'} \big\rangle = - l(l+1)\ \delta_{l,l'} \ \delta_{m,m'}
\end{equation}
The required two-electron repulsion integrals can be found using the standard methods of two-electron integral theory \cite{Review94}.  For example, the spherical harmonic resolution of the Coulomb operator \cite{RO08,Lag09,RRSE09,QuasiRO11,REwald11}
\begin{equation}
	r_{12}^{-1} = R^{-1} \sum_{lm} \frac{4\pi}{2l+1} Y_{lm}^*(\br_1) Y_{lm}(\br_2)
\end{equation}
yields the general formula
\begin{multline}
	\big\langle Y_{l_1m_1} Y_{a_1b_1}\ \big| \ r_{12}^{-1}\ \big| \ Y_{l_2m_2} Y_{a_2b_2} \big\rangle	\\
	= R^{-1} \sum_{lm} \frac{4\pi}{2l+1} \big\langle Y_{l_1m_1}Y_{l_2m_2}Y_{lm} \big\rangle \big\langle Y_{a_1b_1}Y_{a_2b_2}Y_{lm} \big\rangle
\end{multline}
where the one-electron integrals over three spherical harmonics involve Wigner $3j$ symbols \cite{NISTbook} and the sum over $l$ and $m$ is limited by the Clebsch-Gordan selection rules.

%---------------------------------------------------------------------------------
\subsection{Hartree-Fock calculations} \label{subsec:HFcalcs}
%---------------------------------------------------------------------------------
Unlike our calculations for electrons in a cube \cite{Jellium05}, the HF calculations for our present systems are trivial.  Because the shells in $L$-spherium are filled, the restricted\footnote{We note that, in low-density cases, the RHF solutions are unstable with respect to lower-energy, symmetry-broken UHF wavefunctions \cite{TEOAS09}.  However, we will not consider the latter in the present study.} HF and KS orbitals are identical  and are simply the spherical harmonics.  These orbitals yield the reduced energy contributions
\begin{gather}
	\Ts = + \frac{L(L+2)}{4R^2}														\label{eq:Ts}	\\
	\EV = + 0																		\label{eq:EV}	\\
	\EJ = + \frac{(L+1)^2}{R}														\label{eq:EJ}	\\
	\EK = - \frac{L+1}{2R} {_4F_3} \left[	\begin{matrix}
											-L, \quad -1/2, \quad 1/2, \quad L+2	\\
											-L-1/2, \quad 2, \quad L+3/2	
										\end{matrix} \ \ ;\  \ 1 \right]					\label{eq:EK}	\\
	\EX = - \frac{4(L+1)}{3\pi R}														\label{eq:EX}
\end{gather}
where $_4F_3$ is a balanced hypergeometric function \cite{NISTbook} of unit argument\footnote{This hypergeometric may be related to a Wigner $6j$ symbol \cite{NISTbook}.}.  It is encouraging to discover that Eq.~(\ref{eq:EK}) approaches Eq.~(\ref{eq:EX}) in the large-$L$ limit \cite{Glomium11}.

We have used Eqs~(\ref{eq:Ts}) -- (\ref{eq:EK}) to compute the exact HF energies of $L$-spherium for a number of the eigenradii discussed in Section \ref{subsec:quasi}.  These, together with the $\Ts$, $\EJ$ and $\EX$ values, and $\rho$ values from Eq.~(\ref{eq:rho2d}), are shown in Table \ref{tab:results2}.

%-------------------------------------------------------------
\subsection{Orbital-based correlation methods}
%-------------------------------------------------------------
Although it is easy to find the HF energy of $L$-spherium, the calculation of its \emph{exact} energy is not a trivial matter.  How can this best be achieved?  The fact that the occupied and virtual orbitals are simple functions (spherical harmonics), so the AO $\to$ MO integral transformation is unnecessary, suggests that orbital-based correlation methods may be particularly effective.  We now consider some of these.

\paragraph{Configuration interaction \cite{Slater29}}
This was the original scheme for proceeding beyond the HF approximation.  It has fallen out of favor with many quantum chemists, because its size-inconsistency and size-inextensivity seriously hamper its efficacy for computing the energetics of chemical reactions.  Nevertheless, for calculations of the energy of $L$-spherium, its variational character, systematic improvability and lack of convergence issues make it an attractive option.

\paragraph{M{\o}ller-Plesset perturbation theory \cite{MP34}}
In an earlier paper \cite{TEOAS09}, we showed that the MP2, MP3, MP4 and MP5 energies of 0-spherium can be found in closed form, for any value of $R$.  However, although we observed that the MP series seems to converge rapidly for small $R$, its convergence was much less satisfactory for $R \gtrsim 1$, where $\rho \lesssim 0.15$.  Unfortunately, this is useless for our present purposes, because many of the $L$-spherium systems in Table \ref{tab:results2} have much larger radii and lower densities than these values.

\paragraph{Coupled-cluster theory \cite{Bartlett82}}
The coupled-cluster hierarchy (\textit{viz.}~CCSD, CCSDT, CCSDTQ, \ldots) probably converges much better than the MP$n$ series, and this should certainly be explored in the future, but we suspect that it will nonetheless perform poorly in the large-$R$, small-$\rho$ systems where static correlation dominates dynamic correlation \cite{TEOAS09,TwoFaces11,EcBond11} and the single-configuration HF wavefunction is an inadequate starting point.

\paragraph{Explicitly-correlated methods \cite{Kutzelnigg91}}
The CI, MP and CC approaches expand the exact wavefunction as a linear combination of determinants and it has been known since the early days of quantum mechanics that this ansatz struggles to describe the interelectronic cusps \cite{Kato57}.  The R12 methods overcome this deficiency by explicitly including terms that are linear in $r_{ij}$ in the wavefunction, and converge much more rapidly as the basis set is enlarged but, unfortunately, they also require a number of non-standard integrals.  If these integrals are not computed exactly, they are approximated by resolutions in an auxiliary basis set.  In either case, the computer implementation is complicated.

%--------------------------------------------------
\subsection{Other correlation methods}
%--------------------------------------------------
It is possible that the unusually high symmetry of $L$-spherium makes it well suited to correlation methods that are not based on the HF orbitals.  We now consider two of these.

\paragraph{Iterative Complement Interaction (ICI) method \cite{Nakatsuji04,Nakatsuji07}}
In this approach, the Schr\"odinger equation itself is used to generate a large set of $n$-electron functions that are then linearly combined to approximate the true wavefunction.  It has been spectacularly successful in systems with a small number of electrons, but has not yet been applied to systems as large as 2- or 3-spherium (with 18 and 32 electrons, respectively).

\paragraph{Quantum Monte Carlo methods \cite{Luchow00,Foulkes01}}
Of the various methods in this family, Diffusion Monte Carlo (DMC) usually yields the greatest accuracy.  In this approach, the Schr\"odinger equation is transformed into a diffusion equation in imaginary time $\tau$ and, in the limit as $\tau\to\infty$, the ground state energy is approached.  Unfortunately, to be practically feasible, the method normally requires that the wavefunction's nodes be known and, despite some recent progress \cite{Mitas06}, the node problem remains unsolved. The quality of the trial wavefunction and finite-size errors are other potential restrictions \cite{Drummond04,Drummond08}.

%----------------------------------------
\subsection{Results and holes}
%----------------------------------------
The discovery \cite{QuasiExact09} that the Schr\"odinger equation for 0-spherium (and 0-glomium) is exactly solvable for each of its eigenradii is extremely helpful, for it allows us to generate the exact energies $E$ and resulting Kohn-Sham correlation energies $\EcKS$ for the first four atoms in Table \ref{tab:results2}, without needing to perform any of the correlated calculations described above.  However, these are the easiest cases, the ``low-hanging fruit'' so to speak, and there remain large gaps in the Table.  We could have filled these gaps with rough estimates of the exact energies but we prefer to leave them empty, to emphasize that there is much to do in this field and to challenge the correlation experts in the wavefunction and density functional communities to address these beautifully simple systems.

Once this is done, the data in the final three columns of Table \ref{tab:results2} will provide the ingredients for the construction of a new correlation density functional $\EcKS(\rho,\eta)$ that will be exact for all $L$-spherium atoms and for 2D jellium.

Finally, of course, an analogous strategy will yield a new functional that is exact for all $L$-glomium atoms and for 3D jellium.  We will recommend that these functionals replace the LDA correlation functionals that are now being used.

%------------------------------------
\section{Concluding remark}
%------------------------------------
All uniform electron gases are equal, but some are more equal than others.

\begin{acknowledgements}
	We would like to thank the Australian Research Council for funding (Grants DP0771978, DP0984806 and DP1094170) and the National Computational Infrastructure (NCI) for generous supercomputer grants.
\end{acknowledgements}

\end{document}